\documentclass[twocolumn,pre,floats,aps,amsmath,amssymb]{revtex4}
\usepackage{graphicx}
\usepackage{bm}
\usepackage[latin1]{inputenc}
\usepackage{tikz}
\usetikzlibrary{trees}

\usepackage{physics} 

\usepackage{etoolbox}

\makeatletter
\pretocmd{\section}{\addtocontents{toc}{\protect\addvspace{20\p@}}}{}{}
\pretocmd{\subsection}{\addtocontents{toc}{\protect\addvspace{5\p@}}}{}{}
\makeatother

\begin{document}

\title{Optimized protocol to create repeater graph states for all-photonic quantum repeater}
\author{Ming Lai Chan\footnote{mlch@connect.hku.hk}}
\affiliation{Department of Physics, The University of Hong Kong, Pokfulam Road, Hong Kong}
\date{\today}

\begin{abstract}
All-photonic quantum repeater is not yet commercially applicable due to difficulty in the generation of repeater graph states (RGS) and the probabilistic nature of Bell-state measurement. In recent years, several deterministic protocols have been proposed to generate RGS. However, they can only create bare RGS and a rate-distance analysis for the deterministic approach is currently missing. We present a deterministic generation scheme of encoded RGS and show that after optimization, the repeater using our scheme performs at least $4.9$ times better than the traditional probabilistic generation scheme in terms of secret key rate, with a significant reduction in the total number photons by $2$ orders of magnitude. The duration of our generation protocol is only half of the existing deterministic schemes. We also describe in detail an experimental method using cavity QED-enhanced resonance fluorescence and time-bin encoding to implement our protocol. 
\end{abstract}

\maketitle

\section{Introduction}
The physical realization of secure long-distance quantum communication has been the holy grail of quantum information science. Quantum key distribution (QKD) is a scheme that allows two authenticated parties to generate a sequence of shared secret bits called the secret key, which can be used to enable unconditionally secure communication~\cite{QKD}. However, due to exponential fiber attenuation, the rate of secret key generated with direct-transmission QKD protocol $R_{direct}=-\log_{2}(1-e^{-\alpha L})$ suffers from exponential decay~\cite{direct} as the distance $L$ between two parties increases.

To overcome this limitation, two directions of research have been investigated extensively in the past twenty years. The sky-based QKD approach uses a satellite as an intermediate station to distribute secret keys between two remote locations. Since the atmospheric attenuation only has a quadratic scaling in free space, this method has the potential to support high transmission rates. In recent years, tremendous progress has been made in this direction~\cite{satellite,satellite2,satellite3}, yet sky-based QKD has some fundamental constraints like unstable weather conditions~\cite{satellite,satellite3,weather} and the huge construction and maintenance costs for satellites. In addition, for a large-scaled complicated global quantum network in the future,  ultimately we need small scalable ground stations capable of performing quantum cryptographic tasks, which necessitates the need for research in ground-based QKD.

One prominent example of the ground-based QKD is to place many quantum repeater nodes between the sender (Alice) and receiver (Bob). By measuring entangled photons created between nodes, followed by error correction and privacy amplification, a shared secret key can be distilled at a rate $R_{repeater}$ that beats $R_{direct}$. However, building such device often requires matter quantum memories that can entangle with photons and sustain the entanglements before receiving communication signals from other repeater nodes. The all-photonic repeater scheme proposed by Azuma \textit{et al.} can circumvent this requirement by replacing the quantum memory with highly entangled quantum states of photons called the repeater graph states (RGS)~\cite{photonic}. The idea is to create RGS at every other nodes (source nodes) and send photons in the state to the two neighboring nodes (receiver nodes). The receiver nodes then perform probabilistic Bell-state measurements (BSM) on the photons collected. Conditional on the result of BSM, a long sequence of raw bits between Alice and Bob is obtained. While the all-photonic repeater completely eliminates the demand for matter quantum memories, one reason why it is currently not physically realizable is the difficulty in creating RGS. 

Two types of protocols have been proposed to create RGS. One type of RGS generation starts with many pairs of photons and uses fusion gates with measurements to probabilistically fuse them into larger photonic cluster states~\cite{photonic,Mihir}. This approach requires an astronomical number of photons $\sim 10^6-10^9$ per source node even after multiplexing. Another type of generation protocol makes use of quantum emitters and optical pumping to deterministically create RGS via a long pulse sequence~\cite{Buterakos,Antonio,61}. Due to its deterministic nature, the resources overhead of this method is expected to be a lot lower, though a systematic estimation of the overhead is missing in the literature. Moreover, the RGSs created in these deterministic schemes are not equipped with fault-tolerance, meaning if some photons in the bare RGS are lost, the success probability and secret key rate of the repeater would be greatly diminished. To compare with the probabilistic approach, secret key rate of the all-photonic repeater using the deterministic method should also be calculated.

To fill in the above research gaps, we propose a deterministic generation protocol to create encoded RGS. We perform rate-distance analysis~\cite{Mihir} to estimate the secret key rate and resources overhead (total number of photons required, duration of our protocol and the number of gates used) of the repeater with the encoded RGS created using our method. To validate the experimental capabilities of our protocol, we provide an experimental setup and the steps to implement it.

Our protocol is optimized in the sense that the secret key rate is improved with a notable reduction in the total number of photons, time span of the protocol and the number of controlled-phase (CZ) gates required. It has the following seven advantages: (1) it can generate arbitrary encoded RGS; (2) it can be seen as a generalized version of the protocol for small bare RGS in~\cite{Antonio}; (3) our protocol is symmetric, which reduces the generation time of RGS by half compared with the deterministic scheme in~\cite{Buterakos}; (4) we use multiplexing~\cite{Mihir} in our protocol, so the performance of repeater is boosted---in particular, the secret key rate achievable with our repeater is at minimum $4.9$ times higher than the conventional probabilistic protocol~\cite{Mihir}; (5) the total number of photons required is reduced by $2$ orders of magnitude, when compared with the probabilistic scheme in~\cite{Mihir}; (6) the total number of CZ gates is one less than the scheme in~\cite{Buterakos}; and (7) the proposed implementation of our protocol uses picosecond optical pulses for photon generation and hole spin rotation, which can reduce the generation time to nanosecond timescale.

The paper is organized as follows. In Section II, we filter out four construction rules from the literature, which are the necessary ingredients to understand our protocol. In Section III, we briefly review the existing deterministic protocols for the generation of bare repeater graph states (RGS) using quantum emitters and present our own optimized generation protocol for fully encoded RGS with an arbitrary even number of arms. In Section IV, we compute the secret key rate and resource overhead of the repeater using our generation protocol, and compare them with the existing generation protocols. In Section V, we describe a new method to experimentally implement our protocol using cavity-enhanced resonance fluorescence and time-bin encoding. Finally, Section VI provides a discussion and future work.

\section{Construction rules}
Before we describe our generation protocol, it is vital to understand the types of quantum operations performed during the protocol and their effects on cluster states. In this section, we review the quantum operations and condense them into four main construction rules. These rules are the building blocks of RGS generation using quantum emitters, hence are applicable to other solid-state systems including self-assembled quantum dots, nitrogen-vacancy center in diamond and trapped ions. Note that although Refs.~\cite{Buterakos,Antonio} have implicitly used some of the rules, their generation protocols are not optimized in terms of the number of quantum operations, as explained below.
\\
\newline
\noindent \textbf{Rule 1.} \textit{Applying a pumping operation (P) followed by a Hadamard gate (H) on the emitter pushes it away from the emitted photon and creates an entanglement between them.}

Physically speaking, the pumping operation $P$ can be regarded as resonance fluorescence: If an emitter satisfies the double two-level system in Refs.~\cite{machine,Buterakos,Antonio} and is initially in the ground state, exciting the emitter with a linearly polarized light leads to spontaneous emission of a photon with polarization dependent on the level structure. By performing the sequence $H_{d}P$ repeatedly on an emitter $d$ that is first initialized in the $\ket{+}$ state, we can deterministically generate a string of entangled photons. Note that if the emitter belongs to part of a graph state~\cite{Antonio}, we can apply the above sequence again to generate another entangled photon without initializing the emitter. This is seen from the $3$-qubit linear cluster state obtained after two pumping operations with a Hadamard gate applied to the emitter between each pumping. The sequence of operations can be written as $H_{d}PH_{d}PH_{d}\ket{0}_d$. The graphical representation of the sequence is depicted in Fig. 1.
\begin{figure}[ht]
\includegraphics[scale=0.45]{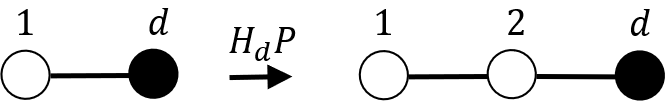} 
\caption{The last generation step of a $3$-qubit linear cluster state. The filled circle represents an emitter. Empty circle represents the emitted photon. Performing a Hadamard gate on the emitter after pumping can be seen as pushing the emitter away from the photons, and creating an entanglement between photon $2$ and emitter $d$. This effect is observed independently in~\cite{Antonio2}.}
\label{fig:4}
\end{figure}
\newline
\noindent \textbf{Rule 2.} \textit{Applying a pumping operation followed by a Hadamard gate on the emitted photon pushes it away from the emitter and creates an entanglement between them.}

In Refs.~\cite{Buterakos,Antonio}, rule 1 is first used. Then two consecutive Hadamard gates are applied to the emitted photon and emitter. However, since the Hadamard gate is an involutory matrix, i.e. $H^2=I$, the generation steps in Fig. 3 of Ref.~\cite{Buterakos}, (a)-(b) and (e)-(f) in Fig. 9 of Ref.~\cite{Antonio} can be reduced to $H_2 H_d H_d P = H_2 P$, which we call rule 2. This rule minimizes the number of Hadamard gates and operation time in deterministic protocols. Similar to the previous example, we create a $3$-qubit linear cluster state using an emitter and optical pumping in Fig. 2, but we apply a Hadamard gate to the emitted photon instead of the emitter. 
\begin{figure}[ht]
\includegraphics[scale=0.45]{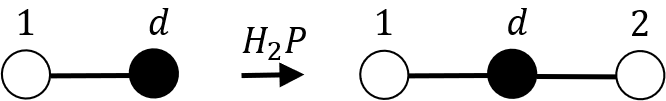} 
\caption{The last generation step of a $3$-qubit linear cluster state with a different pumping sequence. Performing a Hadamard gate on the emitted photon after pumping can be viewed as pushing it away from the emitter.}
\label{fig:5}
\end{figure}
\newpage
\noindent \textbf{Rule 3.} \textit{Performing a controlled-phase (CZ) gate between two coupled emitters creates an entanglement between them.}
\begin{figure}[ht]
\includegraphics[scale=0.4]{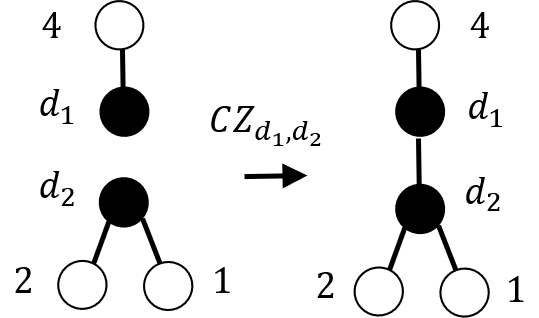} 
\caption{Combine two graph states by applying a CZ gate between two dots.}
\label{fig:6}
\end{figure}

In our protocol, self-assembled quantum dots are chosen. The practical use of CZ gates to create entanglement between tunnel-coupled quantum dots has first been proposed in Ref.~\cite{2D}. The experimental scheme to implement the CZ gate in our protocol is presented in Ref.~\cite{61} and used in Ref.~\cite{Antonio}. Having an operation that entangles both dots is essential in creating two-dimensional RGS, as entangled emitters can produce entangled photons~\cite{2D,Antonio2}. Note that we do not need to initialize both dots with Hadamard gates if they belong to two separate graph states, as shown in Fig. 3. 
\\
\newline
\noindent \textbf{Rule 4.} \textit{Performing local complementation on a qubit, followed by a Z-basis measurement on the same qubit is equivalent to directly applying a X-basis measurement on it.}

Local complementation (LC) is a graph transformation between two local Clifford equivalent states~\cite{X}. Applying Z-basis measurements on a qubit in a graph state removes it from the graph~\cite{Z}. Thus, the position of an emitter in the graph state can be ``controlled", by detaching it from the state with direct Z-basis measurement, then reattaching it to the state with a CZ gate. We use this rule to combine both local complementation and Z-basis measurement into a X-basis measurement, so as to minimize operation time and the number of quantum gates in our protocol. It is discussed in Lemma 1 of Ref.~\cite{X} and used in Ref.~\cite{Antonio}.
\\
\newline
The above four construction rules are used throughout the protocols in Section III. Realistic implementations of these rules are proposed in Section V.

\section{Deterministic generation of RGS}
This section first summarizes the generation protocol of a $4$-armed bare repeater graph state (RGS) in~\cite{Buterakos} and~\cite{Antonio}. A bare RGS is, by definition, a RGS whose inner photons are not encoded. From a practical aspect, a bare RGS is not at all useful because it is not robust against photon loss. In light of this, we present a detailed protocol of the $4$-armed RGS encoded with tree states, and take insights from~\cite{Antonio} to generalize our protocol to create encoded RGS with an arbitrary even number of arms.

\subsection{Generation of a $4$-armed bare RGS}

The generation protocol of a $4$-armed bare RGS is first proposed in Ref.~\cite{Buterakos}. But it is observed independently in~\cite{Antonio} and~\cite{Tzitrin} that some missing entanglements between inner photons at the same repeater node do not affect the functionality of repeater. In fact, the minimum requirement for a usable RGS in all-photonic repeater is that the inner photons must form a complete bipartite graph. Since entanglements between photons are generally difficult to create, it is better off choosing the more simplified and less redundantly entangled RGS. The protocol in Ref.~\cite{Antonio} is summarized in Fig. 4 and explained below.
\begin{figure}[ht]
\includegraphics[scale=0.37]{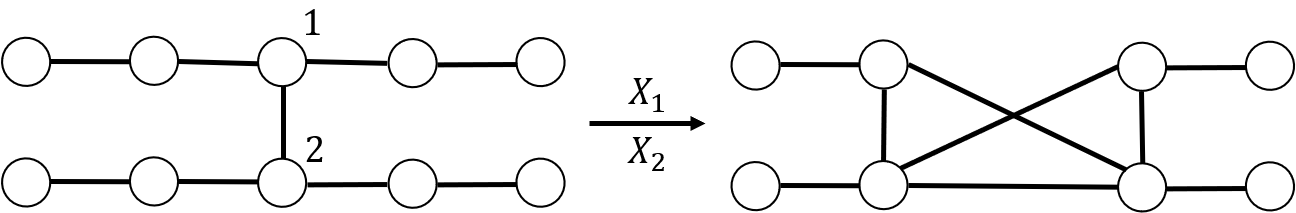}
\caption{Generation protocol of a bare 4-armed RGS suggested in~\cite{Antonio}.}
\label{fig:7}
\end{figure}

We first generate an I-shaped cluster state by applying a controlled-phase (CZ) gate between two tunnel-coupled quantum emitters, where each emitter belongs to a $3$-qubit linear cluster state. Then each emitter is pumped twice under rule 1. Next, we perform Z-basis measurements~\cite{Z} on both emitters to remove them from the state. Using rule 4, two X-basis measurements are performed on qubits $1$ and $2$. Thus, a bare $4$-armed RGS is created. However, if we include photon loss during the generation process and one of the four inner photons in the RGS suffers loss, the performance of repeater will be greatly limited because the loss is equivalent to applying a Z measurement on the missing photon, which is basically the case for unsuccessful Bell-state measurement (BSM)~\cite{photonic}. Thus, a built-in error correction to combat photon loss in RGS is crucial to the repeater. In addition, the above protocol does not generalize to an arbitrary even number of arms, as the steps suggested in Ref.~\cite{Antonio} to create a $6$-arm variant are rather different.
\begin{figure}[ht]
\includegraphics[scale=0.25]{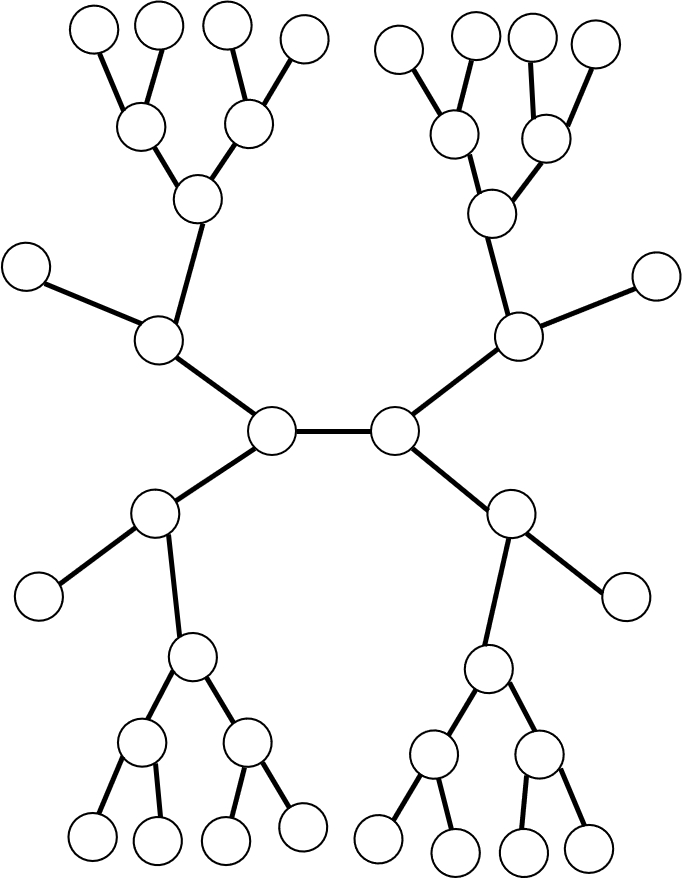} 
\caption{A $4$-armed RGS equipped with $(2,2)$-tree states.}
\label{fig:8}
\end{figure}

\begin{figure*}[ht]
\includegraphics[scale=0.4]{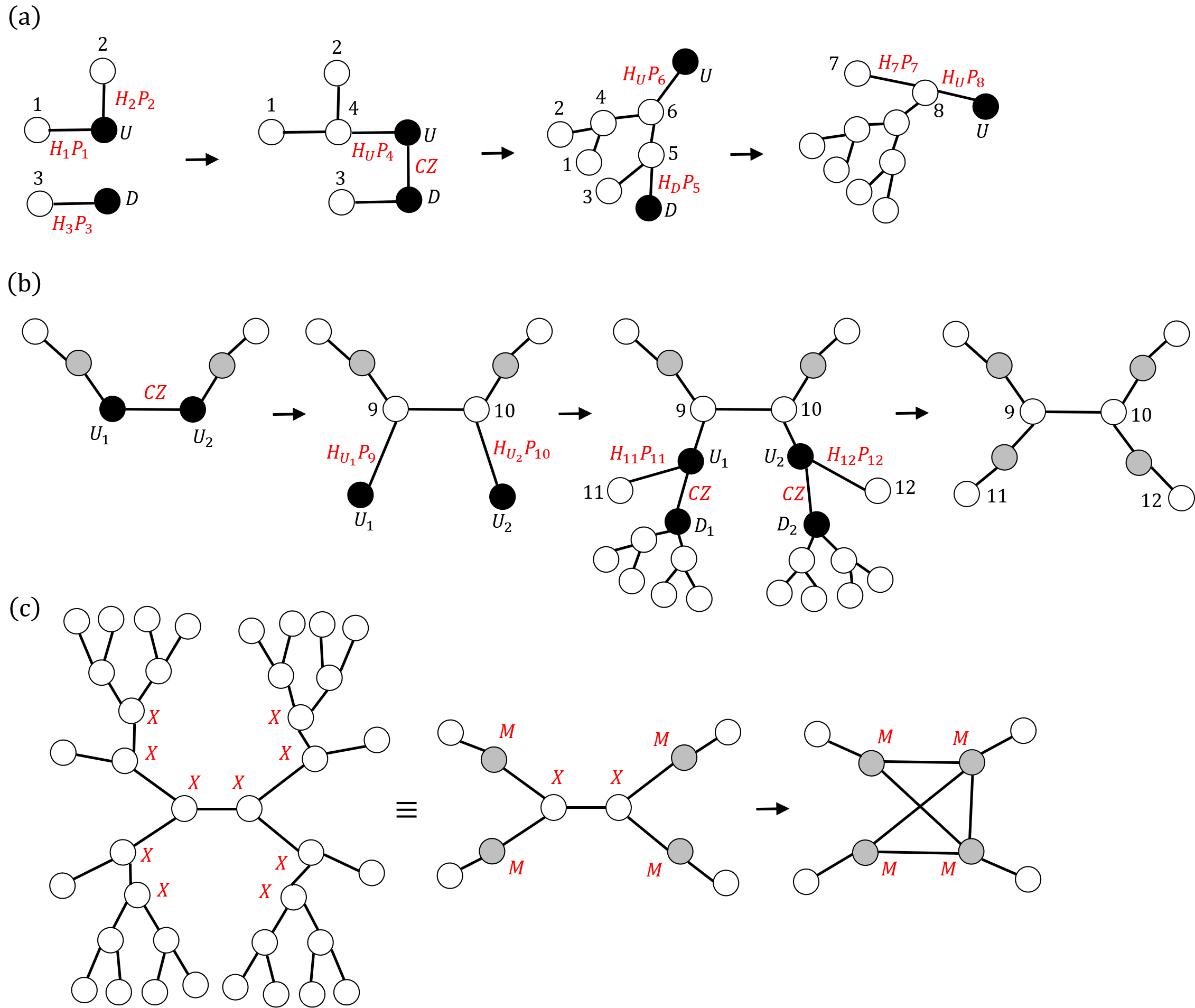} 
\caption{Protocol for the deterministic generation of $4$-armed encoded RGS. Each step is described in the text. (a) First part of the whole protocol. The aim is to obtain two encoded arms by running this sequence in two parallel setups. (b) Second part of the protocol. The goal is to combine the two encoded arms obtained in the first part, and create the RGS in Fig.~\ref{fig:8}. (c) Last part of the protocol. X-basis measurements are performed on the $8$ inner photons and $2$ central photons to complete the encoding. The operations to be performed on the inner photons are denoted as $M$ for brevity.}
\label{fig:9}
\end{figure*}

\subsection{Generation of a $4$-armed encoded RGS}
To make a repeater graph state fault tolerant to photon loss, the method suggested in~\cite{photonic} is to encode the inner photons with tree states. In this section we provide the detailed steps to generate a $4$-armed RGS, where each of four inner photons is connected to a tree state (Fig. 5).

One of the most important considerations in the design of the generation scheme of RGS is the operation time required to generate the RGS, as it indirectly affects the performance of the all-photonic repeater. The protocol in Ref.~\cite{Buterakos} generates each encoded arm in series. This approach is sub-optimal because each iteration of the protocol has time comparable to the spin coherence time of electron when the number of photons of the tree state on each arm is large. Moreover, a common assumption that the protocols in~\cite{Buterakos,Antonio,61} have is, the operation time must be short enough that many iterations can be completed before the electron spin in the quantum dot decoheres. Therefore, the central idea of our protocol is by dividing the generation steps into two simultaneous parallel setups, the total operation time is effectively halved compared with the protocol in Ref.~\cite{Buterakos}, which relaxes the above time requirement. 

Our protocol consists of three parts: (1) each setup generates one encoded arm; (2) connect both arms and generate the remaining arms; and (3) each setup performs X-basis measurements on the inner and central photons to complete the encoded RGS.

In the first part of our protocol, we begin by preparing two vertically stacked and tunnel-coupled quantum dots~\cite{Antonio}. The two quantum dots are labelled according to their positions in Fig.~\ref{fig:9}(a) as $U$ and $D$. Here we follow the same graphical notation in~\cite{Buterakos,Antonio}: Filled circles are quantum dots. Empty circles represent photons. The lines between circles are entanglements. Black texts label the index of a qubit or a dot. Red texts in Fig.~\ref{fig:9}(a) and ~\ref{fig:9}(b) indicate the operations we \textit{have performed}, whereas the red texts in Fig.~\ref{fig:9}(c) represent the operations \textit{to be performed}. Black arrow represents the transition between steps. 

Using rule 2 stated in Section II, we generate photons $1$ and $2$ from dot $U$ and photon $3$ from dot $D$ respectively. Next, we use rule 1 once on dot $U$ to create photon $4$, Subsequently, a CZ gate between both dots is applied to create entanglement.

After the entangling operation, we use rule 1 again to create photon $6$ and push out dot $U$. Simultaneously, rule 1 is performed on dot $D$ to pump out photon $5$. After pumping an additional photon, dot $D$ is then disconnected from the graph, either by a direct Z-basis measurement on the dot, or a Z-basis measurement on another emitted photon~\cite{Antonio}. The last step of first part involves applications of rule 2 to pump out photon $7$ and rule 1 to push out dot $U$.

The tree state in Fig. 5 has depth $2$ and $2$ arms. To deterministically create an arbitrary $(k,d)$-tree state of depth $d$ and $k$ arms, our method uses similar steps to the approach in Ref.~\cite{Buterakos}, except that the root qubit at the $0$-th level in the tree must be a quantum emitter. The steps to generate two such tree states are given and explained in Appendix A. The number of emitters used is equal to the depth of the tree. The number of CZ gates $N_{CZ}$ required in the generation process is therefore slightly different than Eq. (1) in~\cite{Buterakos}, satisfying the revised formula below:
\begin{equation}N_{CZ} = \left\{
\begin{array}{@{}rl@{}}
\frac{k^{d}+(-1)^{d+1}}{k+1}, & \text{even }d;\\
-1+\frac{k^{d}+(-1)^{d+1}}{k+1}, & \text{odd }d.\\
\end{array}
\right .
\label{eq:14}
\end{equation} 
Note that although the above formula only accounts for perfect $k$-ary tree states, our method also works for tree states with arbitrary branching parameters $\{b_i\}_{i=0,1,...,l}$ introduced in~\cite{photonic}, thus the corresponding number of CZ gates can be easily derived. 

Once we obtain two encoded arms from the parallel circuit in the first part of the protocol, we apply a CZ entangling gate between two emitters $U_1$ and $U_2$ and generate the remaining arms. In the second part of the protocol, a grey circle is used to represent the encoded qubit (Fig. 7).

\begin{figure}[ht]
\includegraphics[scale=0.33]{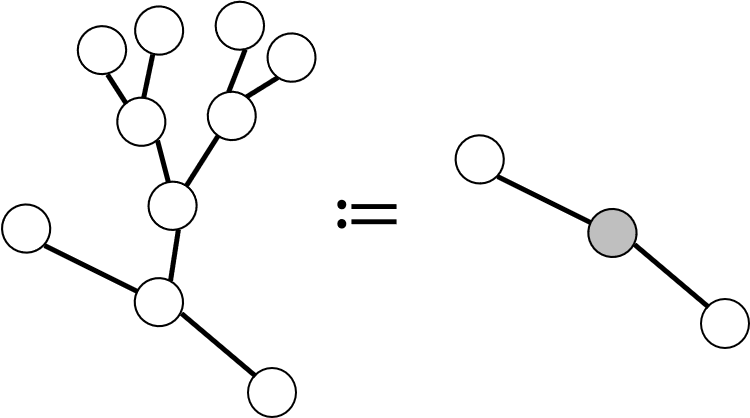} 
\caption{For convenience, we replace the tree state encoded qubit by a grey circle.}
\label{fig:10}
\end{figure}

Consider Fig.~\ref{fig:9}(b). First, the two dots are entangled by a CZ gate using rule 3, so both encoded arms are connected. Then we perform rule 1 once on each dot to pump out photons $9$ and $10$.

To generate the remaining two encoded arms, we run the same steps to create two $(2,2)$-tree states with their root qubits replaced by the dots $D_i$, and attach the trees to the corresponding dot $U_i$ for $i \in \{1,2\}$ under rule 3. Note that we have recycled the dots $D_i$ that were used during first part of the protocol. Now we use rule 2 once on dots $U_i$ to pump out the outer photons $11$ and $12$, and disconnect all the dots from the state. Thus, we obtain the $4$-armed RGS shown in Fig. 5.

For the graph state to be robust against general errors, we follow the idea in~\cite{photonic,Varnava} that X-basis measurements should be applied to the root qubit of the tree and the qubit connecting to it. Essentially, the tree state becomes one single encoded qubit. To complete the encoding in a $4$-armed RGS, we perform $8$ X-basis measurements on the $4$ root qubits as well as their neighboring qubits, where the X-basis measurements on each pair of qubits are defined as $M$.

With the encoding, we can execute loss-tolerant Z-basis and X-basis measurements on the encoded qubits during the later stages of the repeater scheme. The ``magic trick" of tree encoding is when a photon in the encoded qubit is lost, the special quantum correlations on the tree state allows indirect Z-basis measurements, which have the same effect on the measurement outcome as direct Z-basis measurements, as if the photon had not been lost~\cite{Varnava}. In addition, if the local photon loss is less than $50\%$, we can use majority voting to increase the success probabilities of the loss-tolerant measurements~\cite{photonic}, which are critical in the calculations of secret key rate achievable using the repeater, as discussed later in Section IV.

Apart from performing four $M$ operations, we also apply two single-qubit X-basis measurements on the two central photons under rule 4. As a result, the central photons are removed and all inner encoded qubits are connected, reconstructing the same RGS introduced in Fig. 4, whereas the inner qubits are now encoded with $(2,2)$-tree states.

\subsection{Generalized protocol of encoded RGS with $6$ or more arms}
However, a $4$-armed encoded RGS in all-photonic repeater does not offer much advantage over direct repeaterless transmission, as we rely heavily on the success rate of Bell measurements on either of the two qubit channels to obtain inter-node entanglements. In order to fully recognize the effects of encoded tree clusters, we need RGS with higher number of arms. The higher the number of arms, the higher the maximum secret key rate achievable by all-photonic repeater based QKD.
\begin{figure*}[ht]
\includegraphics[scale=0.4]{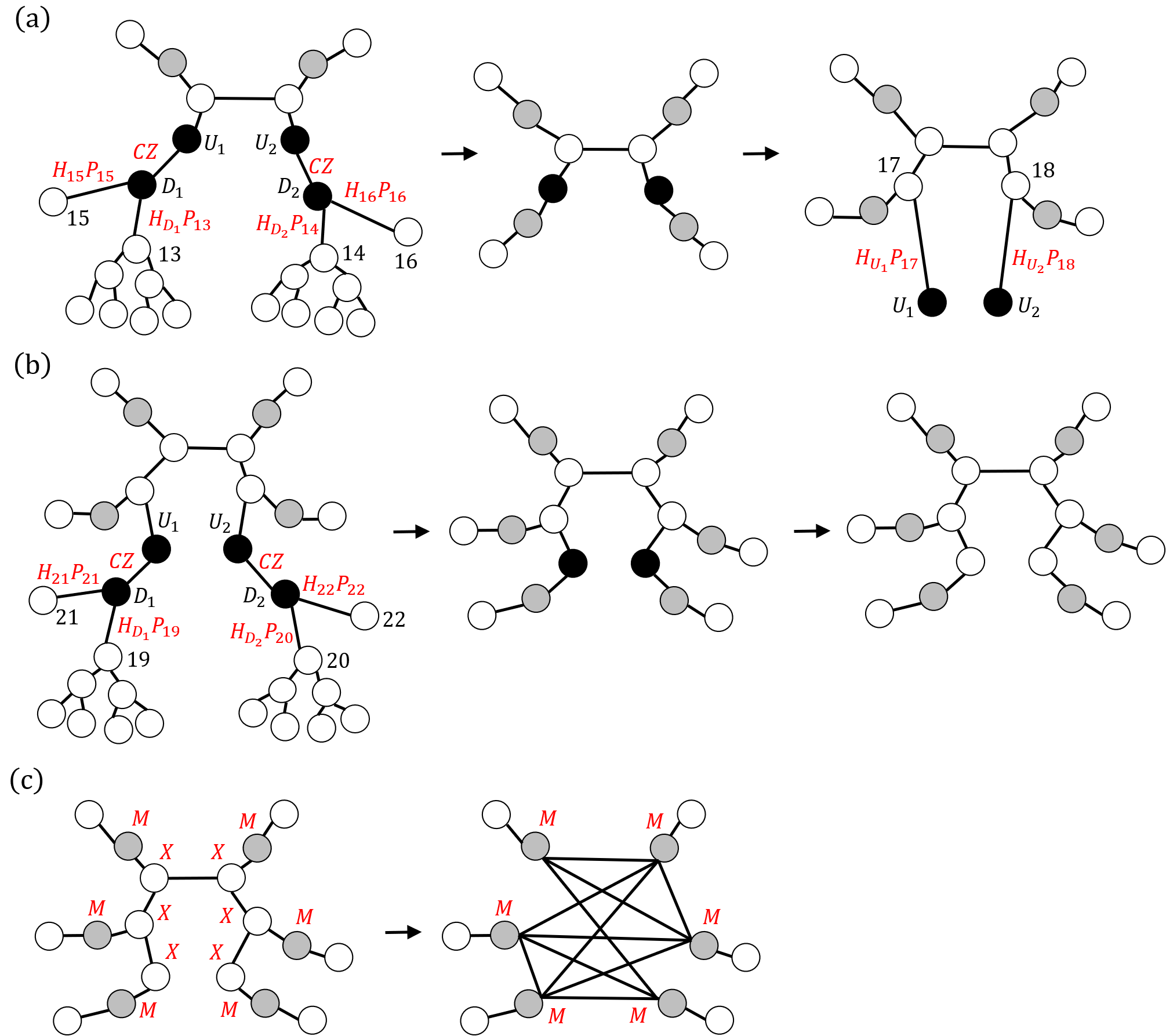} 
\caption{The generalized version of protocol in Fig. 6 for the deterministic generation of encoded RGS. The details are explained in the text. (a) Two extra photons are pumped from each dot $D_i$ before the tree states are connected to the graph. (b) Repeat the sequence in (a) to create two more encoded arms. (c) X-basis measurements are performed on all $12$ inner photons (Denoted by operation $M$), together with $6$ X-basis measurements on the central photons.}
\label{fig:13}
\end{figure*}
Besides, the generation steps suggested in~\cite{Antonio} to create a $4$-arm bare RGS and a $6$-arm variant are fairly different. A more elegant approach would be to take advantage of the symmetry present in the $4$-armed RGS from Fig. 4 and Fig. 6 then generalize to create a RGS of an arbitrary even number of arms. The symmetry argument, together with the potential improvement in secret key rate explain why we need a generalized deterministic generation protocol of RGS.

If we wish to create an encoded RGS with $6$ or more arms, we need to slightly alter the steps in second part of the protocol: We repeat the sequence in Fig. 6(a). After applying a CZ gate to entangle dots $U_i$ and pumping photons $9$ and $10$, we use rule 1 twice on dots $D_i$, followed by rule 2 to pump photons $13$-$16$, as in Fig. 8(a). Then we perform two CZ gates to entangle dots $U_i$ and $D_i$ that have the same index. Lastly, the dots $D_i$ are disconnected and photons $17, 18$ are pumped using rule 1.

Next, we iterate the sequence in Fig. 8(a) to obtain a $6$-armed graph state. Finally, instead of using rule 1 twice on the dots $U_i$, we disconnect them from the state, as depicted in the last step of Fig. 8(b). For $s \geq 1$, we can iterate the sequence in Fig. 8(a) $s+1$ times if we want to create a $(4+2s)$-armed graph state. 

To convert the $6$-armed graph state into a usable RGS, we perform X-basis measurements on the $12$ inner photons and $6$ central photons in Fig. 8(c), similar to the step in Fig. 6(c). Note that if we only hold off all the $M$ operations, the graph state we obtained highly resembles the $6$-armed bare RGS in~\cite{Antonio}, which also has two missing entanglements between qubits that are located on the same side. As discussed previously, the lack of entanglements between qubits on the same side does not affect the functionality of repeater, since the goal of the repeater is to establish connections between outer qubits. Therefore, our protocol is symmetric and can be seen as the generalized version of protocols in~\cite{Antonio}.

The primary reason of having a symmetric protocol is we can divide the generation steps into two simultaneous parallel processes. For example, during the first part of our protocol, we can create two encoded arms at the same time. After applying a CZ gate to entangle the dots $U_i$, the remaining arms in the RGS can also be generated in two parallel setups. Thus, the time span of one run of our protocol is effectively reduced by half compared with the protocol in Ref.~\cite{Buterakos}, which relaxes the requirement in generation time.

\section{Rate Calculations}
In this section, we follow the method used in~\cite{Mihir} to calculate the achievable secret key rate and resources overhead of the all-photonic repeater with RGS created using our deterministic generation protocol described above. We first describe the repeater scheme stated in~\cite{Mihir} after the creation of RGS, followed by an introduction of secret key rates with and without the use of quantum repeater, $R_{repeater}$ and $R_{direct}$. Then we provide an example of RGS to calculate the secret key rates of repeater using probabilistic generation scheme in~\cite{Mihir} and our deterministic generation scheme. Finally, we compare both rates with the rate for direct-transmission QKD $R_{direct}$.

\subsection{Secret key rates with and without repeater}

Our repeater has the same physical structure as the one in~\cite{Mihir}: Alice and Bob are separated by a distance $L$, with source nodes and receiver nodes placed alternatively between them. There are in total $n$ source nodes and $n-1$ receiver nodes. The distance between Alice (Bob) and the nearest source node is $L/2n$, while the distance between two source (receiver) nodes is $L/n$. 

After the repeater graph states are created at each source node, all the encoded photons are stored in a fiber bundle, and the $2m$ outer photons are sent to nearby receiver nodes through optical fibers. As a result, Alice and Bob, who are located at both ends of the repeater, will each receive $m$ outer photons from their closest source nodes. Each receiver node collects $m$ outer photons from a nearby source node. 

In order to create inter-node entanglements between photons, $m$ Bell-state measurements (BSMs) are performed on the $m$ outer photon pairs at each receiver node, while Alice and Bob measure the photons they received in a randomly-chosen basis that they both agreed on. 

The outcomes of BSMs and measurements done by Alice and Bob are then sent to the neighboring source nodes via an authenticated classical channel.

If the BSMs are successful, logical X-basis measurements on the encoded photons will be performed at the source nodes. Otherwise, logical Z-basis measurements will be performed instead. The X-basis measurement has the effect of removing the encoded photons and extending entanglements, whereas Z-basis measurement removes the encoded photons and detaches entanglements.

Whether the repeater scheme is successful in establishing an end-to-end entanglement depends on several conditions: (1) if both Alice and Bob receive a "click" in at least one of their $m$ measurements. The probability of at least one successful detection at each end is denoted by $P_{end}$; (2) if all $n-1$ receiver nodes obtain at least one successful BSMs, with success BSM probability to be $P_B$; and (3) all the logical X-basis and Z-basis measurements on encoded photons, with success probabilities $P_X$ and $P_Z$ respectively, are successful at all $n$ source nodes. 

The conditional probability of obtaining a long-distance entangled photon pair~\cite{Mihir} is therefore,
\begin{equation}
    P_{meas} = P_Z^{2(m-1)n}P_X^{2n}[1-(1-P_B)^m]^{n-1}P_{end}^2.
\end{equation}
If all of the above conditions are met, and after each source node reports its results of logical X-basis and Z-basis measurements to Alice and Bob, a raw bit will be shared between them. Running the repeater multiple times can generate a sequence of raw key, which can be used to distill a shared secret key for QKD to enable secure quantum communication.

Now we find the secret key rate achieved by the repeater, which is defined in~\cite{Mihir} to be the success probability of the whole repeater protocol (which is the joint probability for successful RGS generation at all source nodes with probability $P_{cn}$ and successful long-distance entanglement between both ends with probability $P_{meas}$) divided by the number of spatial channels in each run. Since we assume the use of single-polarization dual-rail encoding, each qubit channel occupies two spatial channels. In total there are $2m$ spatial channels. Hence, the secret key rate $R_{repeater}$~\cite{Mihir} is expressed as
\begin{equation}
\begin{split}
    R_{repeater}&=P_{cn} P_{meas}/2m \\
    &= \frac{P_{cn}}{2m}P_{end}^{2}P_{Z}^{2(m-1)n}P_{X}^{2n}[1-(1-P_{B})^m]^{n-1},
\end{split}
\end{equation}
whereas the secret key rate for repeaterless transmission~\cite{Mihir} is
\begin{equation}
    R_{direct} = -\log_2(1-e^{-\alpha L}),
\end{equation}
with $L$ being the distance (km) between Alice and Bob, and $\alpha$ is the optical fiber loss coefficient.

The probabilities of logical X-basis and Z-basis measurements on the encoded photons discussed above are functions of the success probabilities $\xi_{i}$ of an indirect Z-basis measurement~\cite{photonic,Mihir}, which can be written as:
\begin{equation}
\begin{split}
    P_X &= \xi_0, \\
    P_Z &= (1-\epsilon_{stat}+\epsilon_{stat} \xi_1)^{b_0},
\end{split}
\label{eq:11}
\end{equation}
and the recurrence relation of $\xi_i$ to be
\begin{equation}
    \xi_i = 1-[1-(1-\epsilon_{stat})(1-\epsilon_{stat}+\epsilon_{stat}\xi_{i+2})^{b_{i+1}}]^{b_i},
\label{eq:12}
\end{equation}
with $\xi_{l+1}=\xi_{l+2}=0$, $b_{l+1}=b_{l+2}=0$ and $i \leq l$.

The symbol $\epsilon_{stat}$ represents the loss rate of encoded qubits that are kept locally at the source nodes. It is a function of the survival rate of photons during classical feed-forward steps after measurements on the repeater graph states, thus it is scheme-dependent and varies for different RGS generation protocols.

\subsection{Secret key rates of repeater using probabilistic and deterministic RGS generation schemes}

The loss rate of encoded qubits from the RGS generation scheme using probabilistic fusion gates is found to be~\cite{Mihir}:
\begin{equation}
    \epsilon_{stat} = 1-e^{\frac{-\alpha L}{n}}\eta_{c}\eta_{GHZ}P_{chip}^{k+2}P_{fiber},
\end{equation}
where $\eta_{GHZ}=\eta_{s}\eta_{d}/(2-\eta_{s}\eta_{d})$ is the probability of photons that survive during the creation of GHZ states. $\eta_s$ and $\eta_d$ are the photon source and detector efficiencies. $\eta_c$ is the coupling efficiency between the photonic chip and fiber. It is assumed in~\cite{Mihir} that the repeater graph state is generated on a photonic chip for fast processing and high scalability. $P_{chip} = e^{-\beta \tau_s c_{ch}}$ is the survival rate of a photon on-chip during a feed-forward step after a measurement, say, the X-basis and Z-basis measurements. $\beta$ is the on-chip loss coefficient. $\tau_s$ is the feed-forward time on-chip. $k$ is the minimum number of fusion operations required to create a RGS. $P_{fiber}=e^{-\alpha \tau_{f} c_{f}}$ is the survival probability of photon during a feed-forward step in fiber. The values of all the parameters introduced here are provided in Table. I of Ref.~\cite{Mihir}.

The probability of successful Bell-state measurement using ancilla-assisted fusion gates~\cite{0.75} in the probabilistic generation scheme is
\begin{equation}
    P_{B} = \big[\frac{1}{2}(\eta_{s}\eta_{d})^2+\frac{1}{4}(\eta_{s}\eta_{d})^4\big]e^{\frac{-\alpha L}{n}}(\eta_{c}\eta_{GHZ}P_{chip}^{k+2})^2,
\label{eq:8}
\end{equation}
and the probability of receiving at least one detector click at one's end is
\begin{equation}
    P_{end}=1-\big(1-e^{\frac{-\alpha L}{2n}}\eta_{c}\eta_{GHZ}P_{chip}^{k+2} \big)^m.
\label{eq:9}
\end{equation}
Since only the above three parameters depend on the survival rates of photons during the RGS generation process, we can easily find the expressions of these parameters in other RGS generation protocols and compute their secret key rates.

The loss rate of encoded qubits from our deterministic RGS generation scheme is
\begin{equation}
    \epsilon_{stat}' = 1 - e^{\frac{-\alpha L}{n}}\eta_{c}P_{c1}'P_{chip}P_{fiber},
\label{eq:13}
\end{equation}
where $P_{c1}'$ is the probability of successfully creating a RGS using our generation protocol. We retain the factor $P_{fiber}$ in the expression because the encoded qubits are preserved in a fiber bundle that suffers from the same attenuation as the optical fibers installed between repeater nodes. $P_{chip}$ represents the survival rate of a photon during feed-forward step on-chip after the Bell-state measurement. For our deterministic protocol, the success probability of obtaining a RGS is dependent on the fidelities of all logic gates used, as well as the survival probabilities $P_{chip}$ of photons during feed-forward step after X-basis and Z-basis measurements. We assume the fidelities of single-qubit gates and the single unitary gates on emitters are close to unity, thus $P_{c1}'$ would depend only on the fidelity of CZ gates and $P_{chip}$.

The total number of CZ gates $n_{CZ}$ used in our protocol to generate a RGS of $2m$ ($m \geq 3)$ arms is $2m(N_{CZ}+1)-1$, which is one less than the protocol in~\cite{Buterakos}. As discussed in Section III, $N_{CZ}$ corresponds to the number of CZ gates required to create a tree state. The total number of X-basis and Z-basis measurements performed are $6m$ and $2m+2+2m N_{CZ}$ respectively, since we require $3$ X-basis measurements on each arm to complete the encoding, $2m+2$ direct Z-basis measurements to disconnect the dots in the RGS and $2m N_{CZ}$ more Z-basis measurements to disconnect the dots during the creation process of all tree states. Note that the numbers of CZ gates, X and Z measurements do not follow the above formulas for $m=2$ because we would use the simpler protocol in Fig. 6 instead of the generalized one in Fig. 8.

Multiplexing is a powerful tool used in~\cite{Mihir} that can boost the success probability of RGS creation by simultaneously running through multiple identical generation protocols, with the goal of heralding at least one RGS in $n_{p}$ parallel attempts. The overall success probability $P_{c1}'$ of creating RGS at a source node using our generation protocol and multiplexing is therefore,
\begin{equation}
    P_{c1}' = 1 - (1-F_{CZ}^{n_{CZ}}P_{chip}^{8m+2+2mN_{CZ}})^{n_p},
\end{equation}
with $F_{CZ}$ being the fidelity of a CZ gate, which we take to be $0.995$ using the protocol in~\cite{61}. To achieve this value, three conditions must be met~\cite{Antonio,61}: (1) The spontaneous emission timescale of the quantum dots is a lot less than Larmor frequency and time of exchange interaction. (2) The duration of our generation protocol is short enough that many runs can be completed before the spin in the dot decoheres. (3) The fluctuations in Zeeman frequencies between the dots $U_i$ and $D_i$ should be on the order $10\%$ for equal Zeeman spitting, and order $0.1\%$ for unequal Zeeman splitting. Two methods will be discussed in Section V to partially fulfill the above conditions. The probability that all $n$ source nodes successfully create RGSs is $P_{cn}'=P_{c1}'^n$.

The corresponding probabilities of successful Bell-state measurement and successful detection at one's end are
\begin{equation}
    \begin{split}
        P_{B}' &= \big[\frac{1}{2}(\eta_{s}\eta_{d})^2+\frac{1}{4}(\eta_{s}\eta_{d})^4\big]e^{\frac{-\alpha L}{n}}(\eta_{c} P_{c1}' P_{chip})^2, \\
        P_{end}' &= 1-\big(1-e^{\frac{-\alpha L}{2n}}\eta_{c}P_{c1}' P_{chip} \big)^m,
    \end{split}
\end{equation}
where we have replaced $\eta_{GHZ}P_{chip}^{k+2}$ in Eqs.~(\ref{eq:8}) and~(\ref{eq:9}) by $P_{c1}' P_{chip}$.

\subsection{Scheme comparison}

Now we have all the ingredients to evaluate the secret key rates. Suppose we want to create repeater graph states of $2m=8$ arms with branching vector $\Vec{b}=\{b_0, b_1, b_2\}=\{10,6,3\}$ at $n=314$ source nodes. The total number of photons $Q_l$ in a tree cluster state follows the formula~\cite{photonic}:
\begin{equation}
    Q_l = \sum_{j=0}^{l}\prod_{i=0}^{j}b_{i},
\label{eq:10}
\end{equation}
where $l$ is the largest index in the branching vector $\Vec{b}$. In this example, $l=2$. From the inequality $2m(Q_{l}+1) \leq 2^k +2$ in~\cite{Mihir} and Eq.~(\ref{eq:10}), we know that creating RGS using the probabilistic generation protocol would require a minimum of $k=11$ fusion steps. Similar to the calculations performed in~\cite{Mihir}, we fix $P_{cn}=0.9$ and numerically compute the total number of photons required in the generation process to be in the order of $10^9$.

If we create the RGS using our deterministic generation protocol instead, in order to achieve the same value $P_{cn}'=0.9$, we multiplex the protocol over $n_{p}=89$ parallel attempts. Hence, the total number of photons $N$ required in our protocol is $n_p \times n \times N_{RGS}\approx 5.68\times 10^7$, where $N_{RGS} = 2m(Q_{l}+2)+4m$ is the total number of optical pumping performed to generate a RGS.

To calculate the secret key rates, we first find the expressions for the success probabilities of logical X-basis and Z-basis measurements from Eqs.~(\ref{eq:11}) and~(\ref{eq:12}). We obtain:
\begin{equation}
\begin{split}
    P_X &= 1-[1-(1-\epsilon_{stat})(1-\epsilon_{stat}+\epsilon_{stat}(1-\epsilon_{stat})^{b_2})^{b_1}]^{b_0}, \\
    P_Z &= (1-\epsilon_{stat}+\epsilon_{stat}(1-[1-(1-\epsilon_{stat})^{b_2+1}]^{b_1}))^{b_0}.
\end{split}
\end{equation}
To find the corresponding $P_X$ and $P_Z$ for our deterministic protocol, we can just replace $\epsilon_{stat}$ by Eq.~(\ref{eq:13}). Note that we have chosen a tree depth of $d=3$ for $\Vec{b}=\{b_0, b_1, b_2\}$ because of two reasons: (1) It is shown in~\cite{Mihir} that the performance of repeater improves as $k$ increases. Having trees of larger depth leads to larger $k$. (2) From Eq.~(\ref{eq:14}), we conclude that for odd values of $d$, the number of CZ gates required to create a tree state in our protocol is identical to the one in~\cite{Buterakos}. Choosing a RGS with even $d$ would escalate the total number of CZ gates $n_{CZ}$, thus lowering $P_{c1}'$.
\begin{figure}[ht]
\includegraphics[scale=0.6]{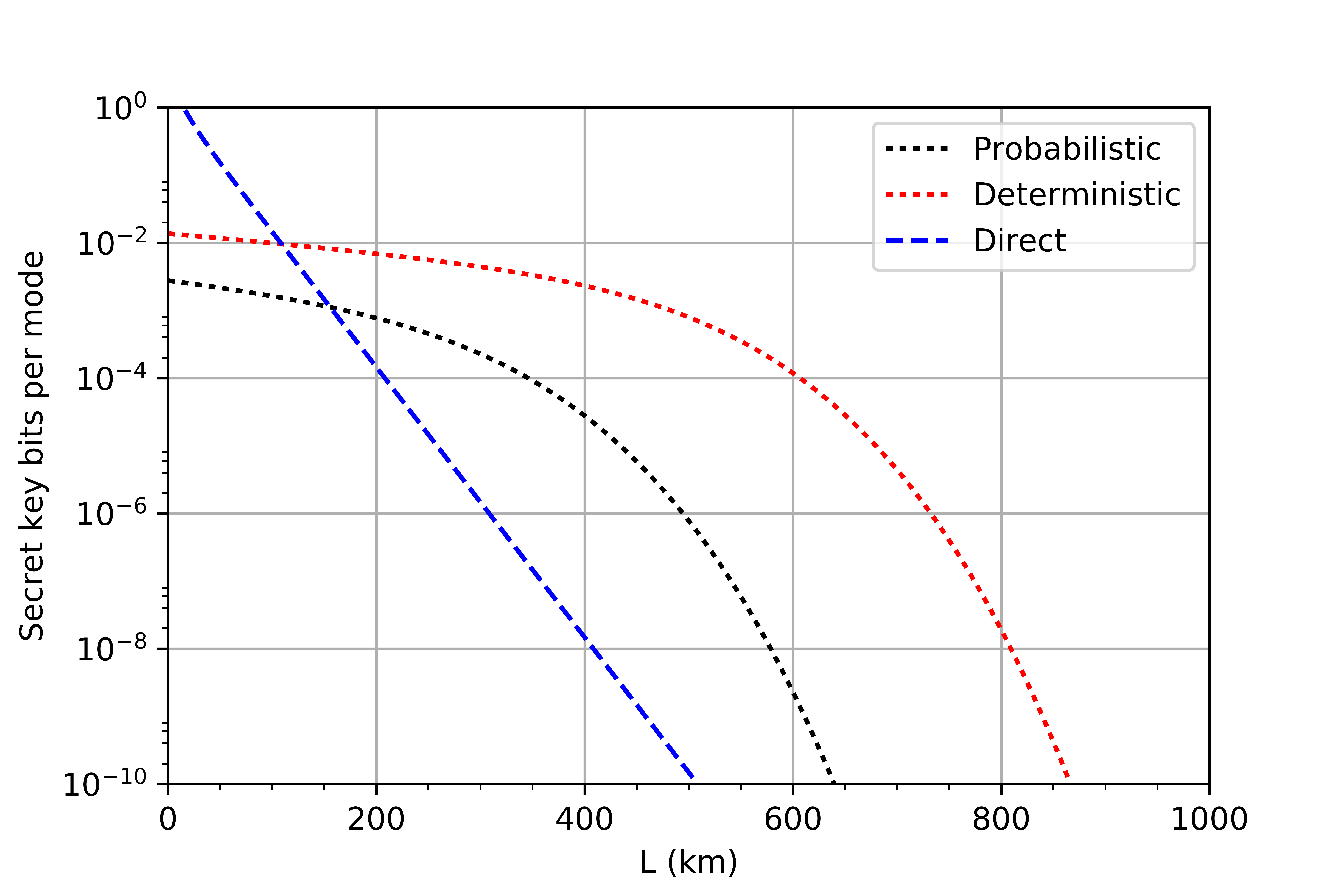} 
\caption{The secret key rates $R_{repeater}$ and $R_{direct}$ are plotted against the end-to-end distance $L$. Black dotted line represents the secret key rate of repeater using probabilistic RGS generation protocol in~\cite{Mihir}. Red dotted line is the key rate of repeater using our deterministic protocol. Blue dashed line is the key rate achieved without the use of quantum repeater. In our example, $n=314$, $m=4$ and $\Vec{b}=\{10,6,3\}$.}
\label{fig:14}
\end{figure}

For better comparison, we make a similar graph to Fig. 6 in Ref.~\cite{Mihir} by plotting the secret key rates $R_{repeater}$ and $R_{direct}$ (bits per mode) as a function of distance $L$ (km).

In Fig.~\ref{fig:14}, it is shown that using our deterministic RGS generation protocol in the all-photonic repeater improves the performance of the repeater, as the secret key rate achievable with our repeater surpasses the rate of repeater using a probabilistic fusion gate approach. The maximum achievable secret key rate of our repeater ($1.37\times 10^{-2}$ bits per mode) is $4.94$ times the rate of the probabilistic approach ($2.77\times 10^{-3}$ bits per mode). As the distance $L$ between Alice and Bob increases, the ratio between the two secret key rates increases exponentially. For $L=600$ km, our repeater performs $5\times 10^5$ times better than the probabilistic protocol. Our repeater exceeds direct-transmission QKD at $109$ km, while the repeater in~\cite{Mihir} requires $156$ km. Using Eq.(18) in Ref.~\cite{Mihir}, the optimal distance between each source node for our protocol is $22.8$ km and $21.5$ km for the probabilistic approach. 

In terms of the resource requirement, the total number of photons used in our repeater is two orders of magnitude less than the probabilistic approach. The duration of one iteration of our RGS generation protocol is halved compared with the protocol in Ref.~\cite{Buterakos}, so the number of possible cycles is twice as many as before. Also, the total number of CZ gates used is $2m\cdot (N_{CZ}+1)-1 = 8(51)-1 = 407$, which is always less than the number of CZ gates in~\cite{Buterakos} by one.

\section{Experimental techniques}
To validate that our theory is compatible with existing experimental capabilities, we will describe a new experimental method to implement our generation protocol in the following section. 

\subsection*{Cavity QED-enhanced resonance fluorescence with the hole spin}

In order to generate repeater graph states, the original approach in Ref.~\cite{Buterakos} relies on the pumping technique suggested in Ref.~\cite{machine}, which makes use of the Faraday magnetic field, enabling a double two-level system for the electron spin. Shining an excitation pulse on the spin after it is initialized in the $(\ket{\uparrow}+\ket{\downarrow})/\sqrt{2}$ state, followed by spin relaxation, will emit photons of circular polarization dependent on the spin state. If we apply a $\pi/2$ rotation pulse on the electron spin afterwards, we would obtain a two-qubit linear cluster state between the electron spin and emitted photon, as described by rule 1 in Section II.

However, the $\pi/2$ spin rotation is performed by applying a weak magnetic field along the Voigt geometry, which activates the diagonal dipole-forbidden transitions and inevitably lowers the quality of the output state. Moreover, the entanglement is carried by the polarization degree of freedom of photons, thus can be destroyed by using polarizers like the dark field microscope in filtering the input and output lasers.

To tackle with the above issues in creating repeater graph states, we use the scheme in Ref.~\cite{Lee} as a basis to realize rules 1-4 introduced in Section II. 

The setup is described as follows: We trap a positively charged quantum dot inside a micropillar cavity to increase photon collection efficiency and place it in a large Voigt magnetic field. The Voigt configuration gives rise to a double-lambda system for the quantum dot consisting of two orthogonal hole states ($\ket{h}, \ket{\Bar{h}}$) and positive trion states ($\ket{T}, \ket{\Bar{T}}$), with the vertical transition $(\ket{\Bar{h}} \to \ket{T})$ being coupled to the cavity. The hole spin is selected over the electron spin as the spin-photon entangler due to its longer spin coherence time ($T_{2}\sim 4 \mu s$)~\cite{coherence}, satisfying condition (2) in Section IV for high fidelity CZ gate. 

Before applying our generation protocol, we need to first initialize the hole spin in the superposition state $(\ket{h}+\ket{\Bar{h}}/\sqrt{2}$. To do so, we use a non-resonant pulse to probabilistically introduce a hole spin into the quantum dot, followed by a resonant pulse to prepare the spin in the $\ket{h}$ state. The two-pulse sequence developed in~\cite{two-pulse} is then used to perform a high fidelity $\pi/2$ spin rotation under the strong magnetic field, resulting in the desired $(\ket{h}+\ket{\Bar{h}}/\sqrt{2}$ state. Note that this two-pulse $\pi/2$ rotation is equivalent to the Hadamard gate ($H$) introduced earlier. 

As for the pumping operation ($P$), it is a combined process of steps 2-4 in the GHZ states generation proposal suggested in Ref.~\cite{Lee}. Therefore, rule 1 in Section II can be realized using the above-mentioned implementations for $H$ and $P$. 

Instead of the polarization mode, the temporal mode of the photon is exploited to be the degree of freedom for generating time-bin entanglement, since photons encoded in the time-bin basis have been shown to be robust against decoherence in fiber transmission~\cite{time-bin}. In Appendix B, we prove that using the above pulses and time-bin encoding, followed by a projective measurement of the hole spin to the $\ket{\Bar{h}}$ state, a two-qubit linear cluster state can be obtained. 

As the sequence in rule 2 involves performing a single-qubit Hadamard gate on a frequency encoded photon, we will use the electro-optic-based frequency beam splitter designed in~\cite{H} to implement this gate. This beam splitter enables a near-unity fidelity ($\sim0.99998$) Hadamard gate that operates in telecom C-band wavelength and can be readily integrated on-chip.

The proposal in Ref.~\cite{61} could be used to implement the CZ entangling gate for rule 3 because of its compatibility with our setup. For example, both setups use the Voigt magnetic field. Moreover, since the quantum dot in our setup is embedded into a microcavity, the decay rate of the transition $\ket{\Bar{h}}\to \ket{T}$ is enhanced by a factor of $\sim5$~\cite{Lee}, which relaxes condition (1) in Section IV.

As discussed in Section II, rule 4 is simply performing a X-basis measurement on a time-bin encoded photon. Time-bin encoding in~\cite{Lee} is achieved by preparing two photon wavepackets that differ in arrival time, where the photon arriving early in an odd number time bin $n$ with state $\ket{e}=\ket{0_{\tau=n+1}1_{\tau=n}}$ is defined to be a logical $1$, and the photon that arrives late with a time delay $\tau_{el}$ in an even number time bin with state $\ket{l}=\ket{1_{\tau=n+1}0_{\tau=n}}$ is defined to be a logical $0$. In order to perform single-qubit measurement in an arbitrary basis, one requires $\tau_{el}$ to be greater than the nanosecond resolution time of photon detector. Otherwise, the late photon will not be measured once it arrives at the detector. 

Ref.~\cite{SFG} has reduced the resolution time to picosecond timescale and experimentally demonstrated ultrafast measurement of a time-bin photon. The idea is to allow oppositely chirped time-bin photon and laser pulse (in a superposition state) to interact in a nonlinear crystal via sum-frequency generation (SFG), outputting a pulse with a spectrum that contains three frequency peaks. The middle peak relates to the probability of successful projective measurement in terms of the relative amplitude $\alpha$ and phase $\beta$, which can be controlled by the shape of laser pulse~\cite{SFG}. Thus, we can use this detector to perform the X-basis measurement in rule 4 by setting $ \alpha=\frac{\pi}{4}$ and $\beta=\{0,\pi\}$ in Ref.~\cite{SFG}.

Therefore, using the above detector for single-qubit X-basis measurement, the several-hundred-picosecond resonant pulse for photon generation and two-pulse sequence for spin rotation~\cite{400ps}, the duration of our generation protocol can be further reduced to the order of several nanoseconds, fulfilling condition (2). 

However, the detector in Ref.~\cite{SFG} comes at a cost of an additional requirement on the time delay $\tau_{el}$: Ideally for high visibility interference in the SFG spectrum, one requires $\tau_{el}$ to be smaller than the temporal bandwidth of the output pulse, and greater than the root-sum-squared coherence time of the two input pulses~\cite{SFG}. It is noted that the time delay in our proposed setup is the time between two photon generation steps. For example, to measure the second photon of the cluster state in Appendix B, $\tau_{el}$, which is the sum of duration of the resonant pulse and two-pulse $\pi/2$ rotation, has to fulfill the above requirement. 

In addition, the non-resonant pulse used to create the hole spin is intrinsically probabilistic, though we argue that this only slightly increases the overhead of our protocol, as we need it once only during the initialization stage---Even if the pulse fails to create the hole spin, we can simply restart the protocol without waste.

\newpage
\section{Discussion}
So far, we have developed a deterministic generation protocol of repeater graph states for all-photonic quantum repeater under the use of four construction rules. Our protocol is optimized in comparison to the existing generation schemes in terms of better performance and lower resources requirement. 

We have performed a rate-distance analysis to compare the secret key rates of repeaters using the probabilistic protocol~\cite{Mihir}, our optimized deterministic protocol and the direct-transmission QKD scheme. We showed that the repeater using our protocol generates a higher secret key rate and requires a lower number of photons than the probabilistic scheme. In particular, the secret key rate generated with our repeater is at least $4.9$ times greater than the rate of repeater using the probabilistic scheme. Our repeater beats direct-transmission QKD at $109$ km, whereas the repeater in~\cite{Mihir} requires $156$ km. 

With multiplexing, our protocol demands $5.68\times 10^7$ number of photons, which is $2$ orders of magnitude less than the one in~\cite{Mihir} ($\sim 10^9$). Due to symmetry in our setup, the total generation time of our protocol is essentially halved compared with the scheme in~\cite{Buterakos}, which doubles the maximum number of cycles that the protocol can run before the spin decoheres. 

To implement our protocol, a new experimental method using cavity-enhanced resonance fluorescence and time-bin encoding has also been thoroughly described, which can resolve the issues arisen from the traditional polarization entanglement scheme in Ref.~\cite{machine}, and possibly lower the total generation time of our protocol to several nanoseconds. As a bonus, the total number of controlled-phase gates required in our protocol is one less than that of the scheme in~\cite{Buterakos}.

Since our generation protocol is built upon the four construction rules given in Section II, the error analysis in~\cite{Buterakos,Antonio,61} also applies to ours. Apart from the existing challenges like low photon extraction efficiency, Pauli errors and spectral inhomogeneity of quantum dots, as well as the time delay requirement introduced in Section V, our protocol requires the loss of each photon to be less than $3$ dB for RGS encoded with tree states to be fault-tolerant~\cite{Varnava,Mihir}. Also, in order to outperform the probabilistic scheme, $89$ parallel attempts of our protocol are needed at each source node, which is still a high resource requirement for practical implementations. 

One interesting future work would be to use the setup and pulse sequence in Section V to experimentally validate rules 1-4 in our protocol. Rules 1 and 2 should be relatively easy to realize, as the operations involved have already been shown feasible in~\cite{Lee,two-pulse,H}. Though, an experimental implementation of the CZ gates used for rule 3 is definitely needed. Successful demonstration of the deterministic generation of small photonic cluster states would not only lead to breakthroughs in the development of quantum repeaters, but also the realization of a scalable quantum network.

\section*{Acknowledgements}
The author would like to thank Zi-Dan Wang, Hoi-Fung Chau, Koji Azuma and Hoi-Kwong Lo for valuable discussions during the early stage of this study, Sen Yang for discussions regarding Raman scattering, Donovan Buterakos for discussions in the deterministic generation of tree states, Mihir Pant for help with the calculation of total number of photons, Martin Hayhurst Appel for stimulating discussions on the generation of time-bin encoded cluster states, and the Quantum Photonics group at the Niels Bohr Institute for warm hospitality. This work was supported by the GRF of Hong Kong (Grants No. HKU173309/16P).

\section*{Appendix A}
As described in Section III, the root qubit at the 0-th level of a tree state used in our repeater graph state (RGS) generation protocol must be a quantum emitter, so that we can attach the tree state to one arm of the RGS via a CZ gate.

In this section, we will present the steps to generate the $(3,2)$- and $(3,3)$-tree states using the construction rules introduced in Section II. The generation steps of an arbitrary $(k,d)$-tree state can henceforth be found.

In Figs. 10-11, red circles represent quantum emitters. Blue circles are photons. Red solid lines are entanglements created by applying a CZ gate between emitters. Blue lines are entanglements created by pumping. Red dash line indicates an emitter is being disconnected by a Z-basis measurement.

Take the $(3,2)$-tree state as an example. As the depth of the tree is 2, we would need 2 emitters to create the whole tree. The idea is to generate one arm using an emitter, then connect the remaining two arms to the emitter using CZ gates.

The procedures to generate the $(3,2)$-tree state are depicted in Fig. 10 and listed below: 
\begin{enumerate}
    \item Create one arm using emitter 1 with rules 1 and 2.
    \item Generate a $(3,1)$-tree with emitter 2.
    \item Apply a CZ gate between two emitters to connect the $(3,1)$-tree to the emitter 1.
    \item Repeat steps 2-3 to attach the remaining arm to emitter 1, thus complete the $(3,2)$-tree.
\end{enumerate}
Note that if we want to create the same tree with the root qubit being a photon instead, one less CZ gate is required because the last arm can be generated using emitter 1~\cite{Buterakos}.
\begin{figure*}[ht]
\includegraphics[scale=0.4]{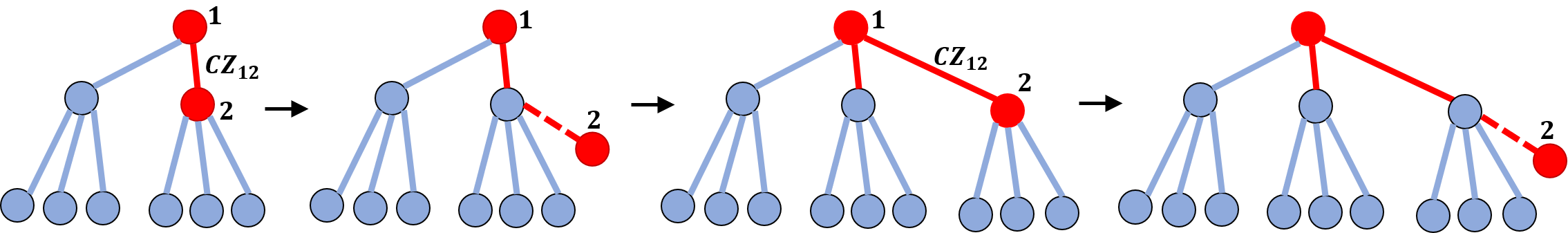}
\caption{Generation of a $(3,2)$-tree state. The number of CZ gates used here is 2, which indeed satisfies Eq. (1).}
\label{fig:15}
\end{figure*}
\begin{figure*}[ht]
\includegraphics[scale=0.4]{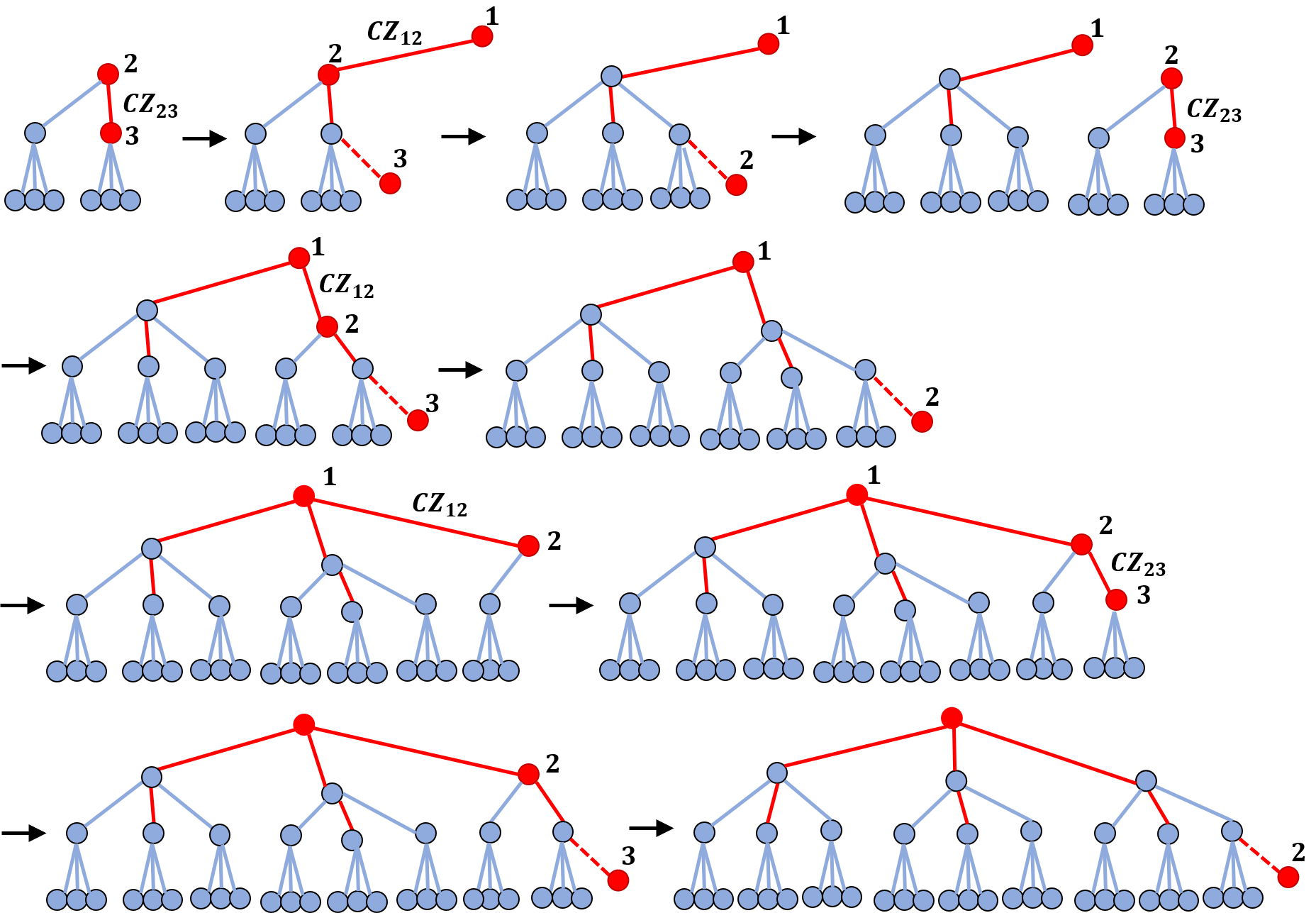}
\caption{Generation of a $(3,3)$-tree state, where each arm consists of a smaller $(3,2)$-tree state. The generation steps can be broken down into creating three $(3,2)$-tree states and connecting them to emitter 1. An arbitrary $(k,d)$-tree state can be recursively generated in a similar manner.}
\label{fig:16}
\end{figure*}
\newpage

For the $(3,3)$-tree state, we need 3 emitters to create the entire tree. From the final tree state obtained in Fig.~\ref{fig:16}, one can quickly observe that emitter 1 is connected to 3 arms, where each arm is a $(3,2)$-tree with an identical distribution of red and blue edges, except that the root qubit on the $(3,2)$-tree is now a photon. The number of CZ gates to create this $(3,2)$-tree is 1 as discussed above. 

Hence, we follow steps 1-3 in the above-mentioned generation protocol of the $(3,2)$-tree state, but with emitters 1 and 2 replaced by emitters 2 and 3. Next, we generate the last arm of the $(3,2)$-tree using emitter 2, and connect it to emitter 1 via a CZ gate. 

At last, by recycling emitters 2 and 3, we can iterate the above procedures twice to create the remaining two arms of the $(3,3)$-tree state.

\section*{Appendix B}
In this section, we will mathematically show that using the proposed implementations for operations $H$ and $P$ in Section V, along with time-bin encoding and projective measurement of the hole spin, a two-qubit photonic cluster state is obtained. Alternatively, one can also view this as a trivial extension of the GHZ state generation protocol in Ref.~\cite{Lee} to cluster state generation.

\subsection{Cluster state generation modified from Ref.~\cite{Lee}}

We consider the same setup in~\cite{Lee} using the Voigt geometry magnetic field and cavity enhanced transition of a positively charged quantum dot. The main difference between generations of the GHZ state and cluster state would be an addition of a $\pi/2$ rotation in between each photon generation step. The steps to generate a two-qubit cluster state are described as follows:
\begin{enumerate}
    \item We follow the same initialization step in Section V and Ref.~\cite{Lee} to obtain a hole spin state $(\ket{h}_s+\ket{\Bar{h}}_s)/\sqrt{2}$. This is equivalent to applying a Hadamard gate on the hole spin with the $\ket{h}$ state, thus we write the sequence of operation as $H_s \ket{h}_s$.
    \item Resonantly drive the cavity enhanced transition $\ket{\Bar{h}}\to \ket{T}$ using a $\pi$-pulse to generate a photon in the first time bin, conditional on the $\ket{\Bar{h}}$ state of the spin. We obtain the state $(\ket{h}_s\ket{0_1}+\ket{\Bar{h}}_s\ket{1_1})/\sqrt{2}$.
    \item Use an off-resonant pulse to perform a spin flip on the trapped hole, resulting in the state $(\ket{\Bar{h}}_s\ket{0_1}+\ket{h}_s\ket{1_1})/\sqrt{2}$.
    \item Resonantly drive the cavity enhanced transition again with a $\pi$-pulse to generate a photon in the second time bin, dependent on the $\ket{\Bar{h}}$ state of the hole spin. We obtain $(\ket{\Bar{h}}_s\ket{1_2 0_1}+\ket{h}_s\ket{0_2 1_1})/\sqrt{2}$. 
    
    Note that if we now use the time-bin encoding, where the state $\ket{1_2 0_1}$ is a logical 0, and $\ket{0_2 1_1}$ as logical 1, it is apparent that steps 2-4 together is the pumping operation $P$: a logical 1 is pumped out when the spin is in $\ket{h}$ state. Otherwise, a logical 0 is pumped out (corresponds to no photon).
    \item Perform a $\pi/2$ rotation on the hole spin using the two-pulse sequence to obtain the state $(\ket{h}_s\ket{1_2 0_1}-\ket{\Bar{h}}_s\ket{1_2 0_1}+\ket{h}_s\ket{0_2 1_1}+\ket{\Bar{h}}_s\ket{0_2 1_1})/2$. This is just another Hadamard gate applied on the hole spin.
    \item Repeating steps 2-5 once leaves us with the state $(\ket{h}_s\ket{1_4 0_3 1_2 0_1}-\ket{\Bar{h}}_s\ket{1_4 0_3 1_2 0_1}-\ket{h}_s\ket{0_4 1_3 1_2 0_1}-\ket{\Bar{h}}_s\ket{0_4 1_3 1_2 0_1}+\ket{h}_s\ket{1_4 0_3 0_2 1_1}+\ket{\Bar{h}}_s\ket{1_4 0_3 0_2 1_1}+\ket{h}_s\ket{0_4 1_3 0_2 1_1}+\ket{\Bar{h}}_s\ket{0_4 1_3 0_2 1_1})/2^{3/2}$. 
    \item Rewriting the above state into $\ket{h}_s(\ket{1010}-\ket{0110}+\ket{1001}+\ket{0101})_{4321}-\ket{\Bar{h}}_s(\ket{1010}+\ket{0110}+\ket{1001}-\ket{0101})_{4321}$ and ignore the coefficients for brevity. We drive the cavity enhanced transition $\ket{\Bar{h}}\to \ket{T}$ and measure the photon generated from the excitation of the enhanced transition, in order to perform a spin readout. After measurement, the quantum state is then projected into $\ket{1010}+\ket{0110}+\ket{1001}-\ket{0101})_{4321}=(\ket{00}+\ket{10}+\ket{01}-\ket{11})_{L}$, which is a two-qubit photonic cluster state under the time-bin encoding $\ket{10}\equiv 0_L$ and $\ket{01}\equiv 1_L$.
\end{enumerate}
The sequence of operation of the two-qubit photonic cluster state is therefore, $M_s H_s P H_s P H_s \ket{h}_s$, where $M_s$ stands for projective measurement of the hole spin. Thus, an $n$-qubit photonic cluster state can be obtained by running the sequence $M_s (H_s P)^n H_s \ket{h}_s$. 

It is important to note that we also need to perform projective measurement of the hole spin during RGS generation discussed in Section V, as two possible RGSs can be obtained depending on the projected spin state. To further elaborate on this point, let us consider cluster state generation. If the hole spin is measured in the $\ket{h}$ $(\ket{\Bar{h}})$ state, the resulting $n$-qubit photonic state is a linear cluster state with the form $S^{(n)}_{+}$ $(S^{(n)}_{-})$ respectively~\cite{Dale}.

\subsection{Proof of equivalence to linear cluster state}
To mathematically prove that the photonic state $S^{(n)}_{+}$ $(S^{(n)}_{-})$ is a linear cluster state, we use the definition and notation in Appendix D of Ref.~\cite{Dale}. For clarity, we will write the pumping operation $P$ as $R F R$ in this section, where $R$ denotes the $\pi$-pulse for photon generation in steps 2 and 4, and $F$ is the off-resonant pulse for spin flip in step 3.

We define the entangled state between the hole spin and $2n$ photons obtained after running our protocol in Section A of Appendix B as $S^{(n)}$, satisfying the relation below:
\begin{equation}
    S^{(n)}=\ket{h}S^{(n)}_{+}-\ket{\Bar{h}}S^{(n)}_{-},
\end{equation}
where $S^{(1)}=\ket{h}(\ket{1_2 0_1}+\ket{0_2 1_1})-\ket{\Bar{h}}(\ket{1_2 0_1}-\ket{0_2 1_1})=\ket{h}S^{(1)}_{+}-\ket{\Bar{h}}S^{(1)}_{-}$ is the state obtained after step 5 of the generation protocol in the previous section. 

Similar to Appendix D of Ref.~\cite{Dale}, we first prove the following lemma by induction:

\textit{Lemma 1}. $\forall n\in \mathbb{N}$, the photonic state $S^{(n)}_{\pm}$ satisfies the recursive relation below:
\begin{equation} \label{eq:16}
    S^{(n)}_{\pm}=\ket{10}_{2n,2n-1}S^{(n-1)}_{+}\mp \ket{01}_{2n,2n-1}S^{(n-1)}_{-}.
\end{equation}
\textit{Proof}. For $n=2$, $S^{(2)}$ is the state obtained after step 6.
\begin{align*}
    S^{(2)} & = HPS^{(1)} \\
    & = HRFR\bigg(\ket{h}S^{(1)}_{+}-\ket{\Bar{h}}S^{(1)}_{-}\bigg)\\
    & = HR\bigg(\ket{\Bar{h}}\ket{0}_{3}S^{(1)}_{+}-\ket{h}\ket{1}_{3}S^{(1)}_{-} \bigg)\\
    & = (\ket{h}-\ket{\Bar{h}})\bigg( \ket{10}_{4,3}S^{(1)}_{+} \bigg)-(\ket{h}+\ket{\Bar{h}})\bigg( \ket{01}_{4,3}S^{(1)}_{-} \bigg)\\
    & = \ket{h}\bigg(\ket{10}_{4,3}S^{(1)}_{+}-\ket{01}_{4,3}S^{(1)}_{-} \bigg) \nonumber \\
    & \quad - \ket{\Bar{h}}\bigg(\ket{10}_{4,3}S^{(1)}_{+}+\ket{01}_{4,3}S^{(1)}_{-} \bigg) \\
    & =\ket{h}S^{(2)}_{+}-\ket{\Bar{h}}S^{(2)}_{-}. \quad \text{From Eq. (15).}\\
\therefore S^{(2)}_{\pm}&=\ket{10}_{4,3}S^{(1)}_{+}\mp \ket{01}_{4,3}S^{(1)}_{-}.
\end{align*}
Assume Eq.~(\ref{eq:16}) is true for $n=k$, where $k \in \mathbb{N}$. Then we consider the case $n=k+1$,
\begin{align*}
    S^{(k+1)} & = HRFR\bigg(\ket{h}S^{(k)}_{+}-\ket{\Bar{h}}S^{(k)}_{-}\bigg)\\
    & = HR\bigg(\ket{\Bar{h}}\ket{0}_{2k+1}S^{(k)}_{+}-\ket{h}\ket{1}_{2k+1}S^{(k)}_{-} \bigg)\\
    & = \ket{h}\bigg(\ket{10}_{2k+2,2k+1}S^{(k)}_{+}-\ket{01}_{2k+2,2k+1}S^{(k)}_{-} \bigg) \nonumber \\
    & \quad - \ket{\Bar{h}}\bigg(\ket{10}_{2k+2,2k+1}S^{(k)}_{+}+\ket{01}_{2k+2,2k+1}S^{(k)}_{-} \bigg) \\
    & =\ket{h}S^{(k+1)}_{+}-\ket{\Bar{h}}S^{(k+1)}_{-}. \quad \text{From Eq. (15).}\\
\therefore S^{(k+1)}_{\pm}&=\ket{10}_{2k+2,2k+1}S^{(k)}_{+}\mp \ket{01}_{2k+2,2k+1}S^{(k)}_{-}.
\end{align*}
For convenience, we use the time-bin encoding discussed in Section V to rewrite Eq.~(\ref{eq:16}) as $S^{(n)}_{\pm}=\ket{0}_{L,n}S^{(n-1)}_{+}\mp \ket{1}_{L,n}S^{(n-1)}_{-}$, where the subscript $L,n$ represents the logical qubit $n$.

If $S^{(n)}_{\pm}$ is an $n$-qubit linear cluster state, it must satisfy the stabilizer condition in~\cite{Dale}:
\begin{equation} \label{eq:17}
    K^{(a)}_{n} S^{(n)}_{\pm} = (-1)^{k^{(a)}_\pm} S^{(n)}_{\pm}, \quad 1\leq a \leq n
\end{equation}
and
\begin{equation*}
    K^{(a)}_{n} = \sigma^{(a)}_{x} \bigotimes_{b \in N(a)} \sigma^{(b)}_{z},
\end{equation*}
where $K^{(a)}_{n}$ is called the stabilizer of state $S^{(n)}_{\pm}$. The superscript $a$ indicates the qubit that the stabilizer is applied on. The value of $k^{(a)}_\pm$ depends on the cluster state we obtained after projective measurement of the hole spin. 

It is important to note that the Pauli operators $\sigma_z$ and $\sigma_x$ are applied on the logical time-bin encoded qubits $\ket{0}_{Ln} \equiv \ket{1_{\tau=2n}0_{\tau=2n-1}}$ and $\ket{1}_{Ln} \equiv \ket{0_{\tau=2n}1_{\tau=2n-1}}$.

Now we can prove the following theorem to show that $S^{(n)}_{\pm}$ is indeed a linear cluster state:

\textit{Theorem 1.} The photonic state $S^{(n)}_{\pm}$ satisfies Eq.~(\ref{eq:17}) with the value of $k^{(a)}_\pm$ to be
\begin{align}
    & k^{(a)}_{+} =
    \begin{cases}
    1, & \text{if }1<a\leq n \\
    0, & \text{if }a=1
    \end{cases} \nonumber \\
    & k^{(a)}_{-} =
    \begin{cases}
    1, & \text{if }1<a< n \\
    0, & \text{if }a \in \{1,n\}.
    \end{cases}
\end{align}

Below we list two lemmas that will be used in the proof. They both hold $\forall n \in \mathbb{N}$ and can be easily proved using induction.

\textit{Lemma 2.}
\begin{equation}
    \sigma^{(n)}_{z} S^{(n)}_{\pm} = S^{(n)}_{\mp}.
\end{equation}
Since the following lemma is not explicitly shown in~\cite{Dale} but is required in the proof, we provide it here.

\textit{Lemma 3.}
\begin{align}
   K^{(a)}_{n+1} = 
   \begin{cases}
   \mathbb{I}^{(n+1)} \otimes K^{(a)}_{n}, & \text{if }a \neq n \\
   \sigma^{(n+1)}_{z} \otimes K^{(a)}_{n}, & \text{if } a=n.
   \end{cases}
\end{align}
where $\mathbb{I}^{(n+1)}$ is an identity operator applied on the $(n+1)$-th logical qubit of the cluster state.

\textit{Proof of Theorem 1.} The base case is trivial so we skip it here. Assume Theorem 1 is true for $n=k$, then we consider $n=k+1$. We separate the proof into four cases: $a=1$, $1<a<n$ for $a \neq n-1$, $1<a<n$ for $a=n-1$ and $a=n$.

\underline{Case 1}: $a=1$. We use Lemma 1 to express $S^{(k+1)}_{\pm}$ in terms of $S^{(k)}_{\pm}$,
\begin{align*}
    K^{(a)}_{k+1} S^{(k+1)}_{\pm} &=K^{(1)}_{k+1} S^{(k+1)}_{\pm} \\
    &= \mathbb{I}^{(k+1)} \otimes K^{(1)}_{k} S^{(k+1)}_{\pm} \\
    &= \mathbb{I}^{(k+1)} \otimes K^{(1)}_{k} \bigg( \ket{0}_{L,k+1}S^{(k)}_{+}\mp \ket{1}_{L,k+1}S^{(k)}_{-} \bigg) \\
    &= \ket{0}_{L,k+1}S^{(k)}_{+}\mp \ket{1}_{L,k+1}S^{(k)}_{-} \\
    &= S^{(k+1)}_{\pm}.
\end{align*}

\underline{Case 2}: $1<a<n$, for $a \neq n-1=k$. Lemma 3 is used to decompose $K^{(a)}_{k+1}$ into $\mathbb{I}^{(k+1)} \otimes K^{(a)}_{k}$,
\begin{align*}
    K^{(a)}_{k+1} S^{(k+1)}_{\pm} &= \mathbb{I}^{(k+1)} \otimes K^{(a)}_{k} \bigg( \ket{0}_{L,k+1}S^{(k)}_{+}\mp \ket{1}_{L,k+1}S^{(k)}_{-} \bigg) \\
    &= -\ket{0}_{L,k+1}S^{(k)}_{+}\pm \ket{1}_{L,k+1}S^{(k)}_{-} \\
    &= -S^{(k+1)}_{\pm}.
\end{align*}

\underline{Case 3}: $1<a<n$, for $a = n-1=k$. Note that since $a=k$, from Eq.~(\ref{eq:17}) we have $K^{(a)}_{k} S^{k}_{+}=-S^{k}_{+}$ and $K^{(a)}_{k} S^{k}_{-}=S^{k}_{-}$, hence
\begin{align*}
    K^{(a)}_{k+1} S^{(k+1)}_{\pm} &= \sigma^{(k+1)}_{z} \otimes K^{(a)}_{k} \bigg( \ket{0}_{L,k+1}S^{(k)}_{+}\mp \ket{1}_{L,k+1}S^{(k)}_{-} \bigg) \\
    &= -\ket{0}_{L,k+1}S^{(k)}_{+}\pm \ket{1}_{L,k+1}S^{(k)}_{-} \\
    &= -S^{(k+1)}_{\pm}.
\end{align*}

\underline{Case 4}: $a=n=k+1$. Lemma 2 is used on the third step.
\begin{align*}
    K^{(a)}_{k+1} S^{(k+1)}_{\pm} &= \sigma^{(k+1)}_{x} \otimes \sigma^{(k)}_{z} \otimes \mathbb{I}^{\otimes_{k-1}} S^{(k+1)}_{\pm} \\
    &= \sigma^{(k+1)}_{x} \otimes \sigma^{(k)}_{z} \otimes \mathbb{I}^{\otimes_{k-1}} \nonumber \\
    & \quad \bigg( \ket{0}_{L,k+1}S^{(k)}_{+}\mp \ket{1}_{L,k+1}S^{(k)}_{-} \bigg) \\
    &= \ket{1}_{L,k+1}S^{(k)}_{-}\mp \ket{0}_{L,k+1}S^{(k)}_{+} \\
    &= \mp S^{(k+1)}_{\pm}.
\end{align*}
Thus completes the proof that $S^{(n)}_{\pm}$ satisfies the stabilizer condition. Therefore, it is a linear cluster state.


\newpage

\begin{thebibliography}{99}  

\bibitem{QKD}Lo, H.-K. and Chau, H.-F., \textit{Unconditional security of quantum key distribution over arbitrarily long distances}, Science \textbf{283}, 2050-2056 (1999).

\bibitem{direct}Pirandola, S., Laurenza, R., Ottaviani, C., and Banchi, L., \textit{Fundamental Limits of Repeaterless Quantum Communications},
Nature Communications \textbf{8}, 15043 (2017).

\bibitem{satellite}Yin. J. et al., \textit{Satellite-based entanglement distribution over 1200 kilometers}, Science \textbf{356}, 1140-1144 (2017).

\bibitem{satellite2}Bedington, R., Arrazola, J.M., and Ling, A., \textit{Progress in satellite quantum key distribution}, NPJ: Quantum Information \textbf{3} 30 (2017).

\bibitem{satellite3}S.-K. Liao et al., \textit{Satellite-to-ground quantum key distribution}, Nature \textbf{549}, 43-47 (2017). 

\bibitem{weather}Hughes, R.J., Nordholt, J.E., Derkacs, D., and Peterson, C.G., \textit{Practical free-space quantum key
distribution over 10 km in daylight and at night}, New J. Phys. \textbf{4}, 43.1-43.14 (2002)

\bibitem{photonic}Azuma, K., Tamaki, K., and Lo, H.-K, \textit{All photonic quantum repeaters}, Nature Communications \textbf{6}, 6787 (2015).

\bibitem{Mihir}Pant, M., Krovi, H., Englund, D., and Guha, S., \textit{Rate-distance tradeoff and resource costs for all-optical quantum repeaters}, Phys. Rev. A \textbf{95}, 012304 (2017).

\bibitem{Buterakos}Buterakos, D., Barnes, E., and Economou, S.E., \textit{Deterministic Generation of All-Photonic Quantum Repeaters from Solid-State Emitters}, Phys. Rev. X \textbf{7}, 041023 (2017).

\bibitem{Antonio}Russo, A., Barnes, E., and Economou, S. E., \textit{Photonic Graph State Generation from Quantum Dots and Color Centers for Quantum Communications}, Phys. Rev. B \textbf{98}, 085303 (2018).

\bibitem{Antonio2}Russo, A., Barnes, E., and Economou, S.E., \textit{Deterministic generation of arbitrary all-photonic graph states from quantum emitters}, (2018), arXiv:1811.06305v1.

\bibitem{61}Gimeno-Segovia, M., Rudolph, T., and Economou, S.E.,
\textit{Deterministic generation of large-scale entangled photonic
cluster state from interacting solid state emitters}, (2018), arXiv:1801.02599.

\bibitem{machine}Lindner, N.H. and Rudoplh, T., \textit{Proposal for Pulsed On Demand Sources of Photonic Cluster State Strings}, Phys. Rev. Lett. \textbf{103}, 113602 (2009).

\bibitem{2D}Economou, S.E., Lindner, N.H., and Rudolph, T., \textit{Optically Generated 2-Dimensional Photonic Cluster State from Coupled Quantum Dots}, Phys. Rev. Lett. \textbf{105}, 093601 (2010).

\bibitem{X}Hahn, F., Pappa, A., and Eisert, J., \textit{Quantum Network Routing and Local Complementation}, (2018), arXiv:1805.04559v2.

\bibitem{Z}Varnava, M., Browne, D.E., and Rudolph, T., \textit{Loss Tolerance in One-way Quantum Computation via Counterfactual Error Correction}, Phys. Rev. Lett. \textbf{97}, 120501 (2006).

\bibitem{Tzitrin}Tzitrin, I., \textit{On the local equivalence of complete bipartite and
repeater graph states}, Phys. Rev. A \textbf{98}, 032305 (2018).

\bibitem{Varnava}Varnava, M., Browne, D.E. and Rudolph, T., \textit{How Good Must Single Photon Sources and Detectors Be for Efficient Linear Optical Quantum Computation?}, Phys. Rev. Lett. \textbf{100}, 060502 (2008).

\bibitem{0.75}Ewert, F. and van Loock, P., \textit{3/4-Efficient Bell Measurement with Passive Linear Optics and Unentangled Ancillae}, Phys. Rev. Lett. \textbf{113}, 140403 (2014).

\bibitem{Lee}Lee, J.P., Villa, B., Bennett, A.J., Stevenson, R.M., Ellis, D. J.P., Farrer, I., Ritchie, D.A., and Shields, A.J., \textit{Towards a source of multi-photon entangled states for linear optical quantum computing}, (2018), arXiv:1804.11311v1.

\bibitem{coherence}Huthmacher, L., Stockill, R., Clarke, E., Hugues, M., Le Gall, C., and Atatüre, M., \textit{Coherence of a dynamically decoupled quantum-dot hole spin}, Phys. Rev. B \textbf{97}, 241413(R) (2018).

\bibitem{two-pulse}Mizrahi, J., Neyenhuis, B., Johnson, K.G., Campbell, W.C., Senko, C., Hayes, D., and Monroe, C., \textit{Quantum control of qubits and atomic motion using ultrafast laser pulses}, Applied Physics
B \textbf{114}, 45 (2014).

\bibitem{time-bin}Marcikic, I., de Riedmatten, H., Tittel, W., Zbinden, H., Legre, M., and Gisin, N., \textit{Distribution of Time-Bin Entangled Qubits over 50 km of Optical Fiber}, Phys. Rev. Lett. \textbf{93}, 180502 (2004).

\bibitem{H}Lu, H.H., Lukens, J.M., Peters, N.A., Odele, O.D., Leaird, D.E., Weiner, A.M., and Lougovski, P., \textit{Electro-optic frequency beam splitters and tritters for high-fidelity photonic quantum information processing}, Phys. Rev. Lett. \textbf{120}, 030502 (2018).

\bibitem{SFG}Donohue, J.M., Agnew, M., Lavoie, J., and Resch, K.J., \textit{Coherent Ultrafast Measurement of Time-Bin Encoded Photons}, Phys. Rev. Lett. \textbf{111}, 153602 (2013).

\bibitem{400ps}Lee, J.P., Wells, L.M., Villa, B., Kalliakos, S., Stevenson, R.M., Ellis, D.J.P., Farrer, I., Ritchie, D.A., Bennett, A.J. and Shields, A.J., \textit{Controllable Photonic Time-Bin Qubits from a Quantum Dot}, Phys. Rev. X \textbf{8}, 021078 (2018).

\bibitem{Dale}Scerri, D., Malein, R. N. E., Gerardot, B. D., and Gauger, E. M., \textit{Frequency-encoded linear cluster states with coherent Raman photons}, Phys. Rev. A \textbf{98}, 022318 (2018).

\end{thebibliography}
\end{document}